\documentstyle[12pt]{article}
\begin{document}
\begin{center}
\Large{Non-equilibrium relaxation and interface energy of the Ising model}

\bigskip

\large{Nobuyasu Ito}

\medskip

\large{Institute for Theoretical Physics\\ University of Cologne\\
       D-W-5000 Cologne 41, Germany \\

and\\

Computing and Information Systems Center \\
Japan Atomic Energy Research Institute \\
Tokai, Ibaraki 319-11, Japan \\
}

\end{center}

\bigskip

\noindent {\bf Abstract}

\smallskip

{}From the non-equilibrium critical relaxation study of the
two-dimensional Ising model, the dynamical critical exponent $z$ is estimated
to be $2.165 \pm 0.010$ for this model.
The relaxation in the ordered phase of this model is consistent with
$\exp (-\sqrt{t/\tau })$ behavior.
The interface energy of the three-dimensional Ising model
is studied and the critical exponent
of the correlation length $\nu$ and the critical amplitude of the surface
tension $\sigma_0$ are estimated to be $0.6250\pm 0.025$ and
$1.42\pm 0.04$, respectively.
A dynamic Monte Carlo renormalization group method is applied to the
equilibrium properties of the three-dimensional Ising model successfully.

\vfill

\pagebreak

\section{Introduction}

The study of the non-equilibrium relaxation turned out to be useful to
study the critical-slowing down and other dynamical aspect of the
Ising models\cite{KO86,DS92,DS92B,KS92,NI93,MH93}.

For the critical relaxation, it is shown that the analysis
based on the dynamical finite-size scaling theory\cite{MS76} produces
precise estimation of the value of $\beta /z\nu$, where $\beta$ and $\nu$
denote the equilibrium critical exponents of the spontaneous magnetization
and the
correlation length, respectively, and $z$ denotes the dynamical critical
exponent.
It is estimated directly from the initial relaxation curve of the
magnetization $m(t)$ which follows asymptotically $t^{-\beta /z\nu}$.
This method is simpler than
the other previous methods which require the calculations of the
two-time correlations and the estimations of the correlation times from them.
In this paper, the values of $z$ for the two-dimensional Ising model
is studied from the non-equilibrium relaxation.
The exact values of $\beta$ and $\nu$ of this model are known to be $1/8$
and $1$, respectively, and the value of $z$ is estimated from the value of
$\beta /z\nu$.

The non-equilibrium relaxation  analysis is useful also for the
relaxation process in the ordered phase which is not yet fully understood.
It has been pointed out that the equilibrium relaxation
of the magnetization may be stretched exponential, that is,
\begin{equation}
C_m(t)\sim {\rm e}^{-at^\alpha},
\end{equation}
where $C_m(t)$ denotes the equilibrium time-correlation function
of the magnetization and $\alpha \leq 1$.
Two phenomenologies predicts $\alpha =1/2$\cite{HF87} and $1/3$\cite{TNM88}
for the two-dimensional Ising model.
In this paper, the non-equilibrium relaxation in the ordered phase is
studied assuming the similar stretched exponential decay, that is,
\begin{equation}
m(t) - m_s \sim a{\rm e}^{-bt^\alpha }.
\end{equation}

The value of $\beta /z\nu$ for the three-dimensional Ising model was
estimated to be $0.250\pm 0.002$\cite{NI93}.
The value of $\beta /\nu$ of this model has been estimated by
several methods and the non-equilibrium critical relaxation
provides one method to study its value.
When the scale transformation with scale $l$ is applied $n$ times
repetitively, the scale transformed magnetization is expected to
follow the scaling form
\begin{equation}
m(t,n) = l^{n\beta /\nu} f(tl^{-nz})
\end{equation}
at the critical point for large $tl^{-nz}$.

In this scaling region, the magnetization at the fixed time
follows\cite{NIMS87,NIMS88}
\begin{equation}
m \sim l^{n\beta/\nu}.
\end{equation}
When we use the total magnetization $M$ instead of the magnetization
per site, it behaves as
\begin{equation}
M \sim l^{-n(d-\beta /\nu )}.
\label{EQFRACT}
\end{equation}
This relation provides a new method of Monte Carlo renormalization group.
The study of $\beta/\nu$ for three-dimensional Ising model based on this
scaling argument is also included.

The second problem of this paper, the interface energy and the
surface tension\cite{BW72}.
In the recent studies of this problem\cite{MH92,KM92,GM92,BHN92,HP93,GPRS93},
it seems to be believed that the standard Monte Carlo simulation with
single spin updating dynamics is not an appropriate method for this
problem.
In this paper, however, it is shown that the naive simulation
works successfully and a precise study of the interface energy is made.
We reconfirm
the values of the critical exponent and the critical amplitude, which is
important to verify the universality hypothesis
about the critical amplitude\cite{KB83}.

The non-equilibrium relaxation study of the two-dimensional Ising model
is described in the next
section and the interface energy of the three-dimensional Ising model
in the third section.
The $\beta /\nu$ analysis is given in the appendix~A.
The last section contains the summary.

\section{Two-dimensional Ising model}

The non-equilibrium relaxation phenomena is studied for the
two-dimensional Ising model in this section.

\subsection{Dynamics and algorithm}

The Hamiltonian of the ferromagnetic Ising model on square lattices is
\begin{equation}
-\beta H(\{\sigma \} )=K\sum_{|i-j|=1} \sigma_i \sigma_j \quad (K>0, \sigma_i
=\pm 1),
\end{equation}
where $\beta =1/kT$ and the summation runs over all the nearest neighbor
site pairs.
The dimensionless constant $K$ is used as the inverse temperature in
the following.
The Ising spins at the lattice points are denoted by
\begin{equation}
\sigma_{(i,j)}\quad (i=1,2,\cdots, L_x, j=1,2,\cdots, L_y),
\end{equation}
where $L_x$ and $L_y$ are assumed to be $L+1$ and $L$, respectively,
and $L$ is an even integer.
Skew boundary condition is applied to $L_x$ direction and
periodic boundary condition to $L_y$ direction.
There is no intrisic dynamics of this model and the dynamics studied here
is the single-spin updating stochastic dynamics with
two-sublattice flip sequence and Metropolis-type transition probability.
The spins are classified into two sublattices defined by
\begin{equation}
\Omega_A =\{ \sigma_{(i,j)}; i+j =\mbox{ even } \}
\mbox{\ and\ }
\Omega_B =\{ \sigma_{(i,j)}; i+j =\mbox{ odd } \} .
\end{equation}
The spins in $\Omega_A$ are updated firstly and then those in $\Omega_B$
are updated.
The transition probability is
\begin{equation}
p_{\rm M}(\sigma_0 \rightarrow -\sigma_0;
  \sigma_1, \sigma_2, \sigma_3, \sigma_4,
) = \mbox{min}
\{ 1, \exp [-2E(\sigma_0; \sigma_1, \sigma_2, \sigma_3, \sigma_4 ) ]\} ,
\end{equation}
where $\sigma_1$, $\sigma_2$, $\sigma_3$ and $\sigma_4$ are the
nearest-neighbor spins of $\sigma_0$ and the local energy $E$ denotes
\begin{equation}
E(\sigma_0; \sigma_1, \sigma_2, \sigma_3, \sigma_4 )
=
K\sigma_0 (\sigma_1 +\sigma_2 +\sigma_3+\sigma_4 ).
\end{equation}

This dynamics is efficiently vectorized and it is the reason why it is
used.
$64$ independent systems with the same lattice are simulated simultaneously.
The $64$ spins on the same site of these $64$ systems are coded in the same
eight-byte integer variables, that is, the independent-system coding
technique is employed\cite{IK88,IK90A,KIK92}.
The spin-updating procedure is realized by $15$ logical operations and
the generation of one random integer.
The condition that a spin $\sigma_0$ is flipped to $-\sigma_0$ is
\begin{equation}
r
\leq
p_{\rm M}(\sigma_0, -\sigma_0 ; \sigma_1, \sigma_2, \sigma_3, \sigma_4 ),
\end{equation}
where $r$ is a random number.
The transition probability $p_{\rm M}$ is the function of local energy
$E(\sigma_0; \sigma_1, \sigma_2, \sigma_3, \sigma_4 )$.
The spin variable $\sigma_i$ which takes $-1$ or $+1$ value is stored in
one bit of one computer word which takes $0$ or $1$ value and the
corresponding bit is denoted by $I_i$.
Now the above condition is expressed by
\begin{equation}
I_1\oplus I_0 +
I_2\oplus I_0 +
I_3\oplus I_0 +
I_4\oplus I_0 +
2*IX1T(r)+IX2T(r) \geq 2 ,
\end{equation}
where $\oplus$ denotes the exclusive-OR operation and
$IX1T(r)$ and $IX2T(r)$ are the functions of random number $r$ defined
by
\begin{equation}
IX1T(r) =\lbrace
\begin{array}{cc}
1 & r\leq {\rm e}^{-8K} \\
0 & \mbox{ otherwise } \\
\end{array}
\end{equation}
and
\begin{equation}
IX2T(r) =\lbrace
\begin{array}{cc}
1 & {\rm e}^{-8K}< r\leq {\rm e}^{-4K} \\
0 & \mbox{ otherwise } \\
\end{array}.
\end{equation}
This condition should be checked by the bitwise logical operations.
The following operations realize it with 14 logical operations.
In the following, $<I,J>$ denotes $2*I+J$.
\begin{enumerate}
\item{}Make $J1$ and $J2$ so that $<J1,J2>=I1+I2+I3$.
It takes five operations.
\item{}Take the exclusive-OR of $J1$, $J2$ and $J4$ with $I0$, respectively.
It takes three operations.
The same notations are used for the results.
\item{}Make $K1$ and $K2$ so that $<K1,K2>=I4+IX1T(r)+IX2T(r)$.
It takes three operations.
\item{}Make $ID$ which holds the truth value of spin-flip condition by
\begin{equation}
K1 \mbox{ or } J1 \mbox{ or } (J2 \mbox{ and } K2).
\end{equation}
It takes three operations.
\end{enumerate}
Further details of this updating algorithm are found in the
FORTRAN routines given in the Appendix~B.
To do the independent simulation of the systems at the same temperature
with the independent-system coding technique, the recycle algorithm of the
random numbers is used by shuffling the Boltzmann factor tables\cite{IKO93}.
The RNDTIK routine\cite{IK90B} are used for the random number generations.

The NEC SX3/11 of K\"oln university and the MONTE-4 of Japan Atomic
Energy Research Institute are used for the simulation.
The performance of the simulation codes developed for the present study,
named IS2DVP, on these vector processors is shown in Table~\ref{PFIS2DVP}.
The MONTE-4 has four processors and they share the main memory.
So the simulation speed on one processor is influenced by other jobs on other
processors.
The performance in Table~\ref{PFIS2DVP} is measured with/without other
jobs on other processors.
The concurrent jobs were Monte Carlo simulations of two- and
three-dimensional Ising models which are highly vectorized.
The speed-down shown in Table~\ref{PFIS2DVP} under multi-job environment can
be regarded as the worst case.
If the concurrent jobs are badly vectorized jobs or the jobs which are
highly vectorized but do not generate the main memory access massively,
the speed-down becomes smaller.
In fact, almost no speed-down was observed when the performance of
IS2DVP is measured with three lowly vectorized jobs.

The value of magnetization at $t$MCS from all up initial condition
is denoted by $m(t)$.
This time-dependent magnetization of the above-mentioned dynamics
is estimated by the simulation.
It is estimated from the values of magnetization at $t$
of independent Monte Carlo runs.

\subsection{At the critical point}

The critical point of the square-lattice Ising model is
$K_c = J/k_{\rm B}T_c=$ $0.440686793\cdots$ where the
time-dependent magnetization
$m(t)$ decays to zero following asymptotically to
\begin{equation}
m(t) \sim t^{-\beta/z\nu} = t^{-1/8z},
\end{equation}
where the exact values of the critical exponents of the spontaneous
magnetization and the correlation length, $\beta =1/8$ and $\nu =1$, are used.

The simulations are listed in Table~\ref{KCSIMUL}.
The deviation of $m(t)$ due to the finite lattice is expected
to be exponentially small for large system.
The system-size dependence at several time
are shown in Fig.~\ref{CRITSIZEDEP}.
No size dependence is observed up to $t=400$MCS, although
it is observed at $t=500$MCS.
In the following, the estimated values of $m(t)$ for $L=1500$ lattice
up to $400$MCS which are shown in Fig,~\ref{CRITMAGNET}
are used for estimating the value of $z$.

Firstly the estimator
\begin{equation}
R(t, m) = t\lbrack {m(t-1) \over m(t)} -1\rbrack
\end{equation}
is used for this purpose.
The average of this $R(t)$ over $N$ successive $t$, that is,
\begin{equation}
R_{\mbox{ave.}}^N (\tau (t,N), m)
  = {1\over N} \sum_{\tau =t}^{t+N-1} R(\tau ,m),
\end{equation}
is used to reduce the statistical fluctuation,
where the mean time, $\tau (t,N)=t+(N-1)/2$.
The values of $z_{\mbox{ave.}}^{25}=1/(8R^{25}_{\mbox{ave.}})$
is plotted in Fig.~\ref{CRITRATIO}.
If the form of $m(t)$ is $t^{-\lambda}(a_0+a_1/t+a_2/t^2+\cdots )$,
$R^N_{\mbox{ave.}}(\tau , m(t))$ is the form of
$\lambda + d_1/\tau +d_2/\tau^2 +\cdots $.
It is observed that the value of $z$ converges to $2.165 \pm 0.005$
assuming that the correction to the asymptotic simple-power law is analytic,
which is consistent with our data.

The estimates for $z$ using the least-square fitting analysis are shown
in Fig.~\ref{CRITFITTING}.
The function of the form of $\log m(t) = a-b\log t$ is fitted to the
successive fifty values of magnetization, that is,
$m(t)$, $m(t+1)$, $\cdots$ $m(t+48)$ and $m(t+49)$.
The values of $z^{50}_{\mbox{fitting}}=8/b$ are plotted.
In the large time limit, the value of $z$
estimated from this Fig.~\ref{CRITFITTING} is consistent with the estimated
$z=2.165\pm 0.005$ from the ratio analysis.

There are many theoretical studies on this value of
$z$\cite{KK72,HH77,MV81,LTW88,LVG93}.
The estimates for the two-dimensional Ising model are listed in
Table~\ref{CRITZEST} and they are plotted in Fig.~\ref{CRITZPLOT}.
It is observed that the simulational values which includes the
estimates obtained by temperature dependence analysis, finite-size scaling
analysis, Monte Carlo renormalization group analysis, non-equilibrium
relaxation analysis and other analysis of the simulational results are
converging.

Above analysis assumes the form of the function $m(t)$ to be a simple
power-law with simple correction terms and the results might have
unknown systematic error caused by this assumption.
Therefore the naive error region might be underestimated and an
overestimated error region will be safer as our present conclusion.
So we adopt $z=2.165 \pm 0.010$ as our conclusion.

There is a room for the qualitatively different assumption for the
critical-slowing down phenomena from above-mentioned simple power law.
For example, a behavior with logarithmic correction to the naive diffusive
behavior, $z=2$, was proposed\cite{OTHER84A}.
Assuming $m(t)$ to be $t^{-\beta /z\nu} (\log t)^{-c}$,
where $\beta /z\nu = 1/16$, the values of $t^{1/16}m(t)$ is plotted in
Fig.~\ref{CRITLOGLOG}.
The value of $c$ is roughtly estimated to be $0.04$ from the tangent
of the plot about $t=200 \sim 400$($\log t = 5.3\sim 6.0$), although
the convex curvature seems to remain.

\subsection{In the ordered phase}

The simulations listed in Table~\ref{LOWMCS} are made.
Each run started from all up initial configuration.
In the previous section, we observed that the value of the magnetization
of $L=1500$ lattice up to  $400$MCS at the critical point can be
regarded as values in the thermodynamics limit with the present accuracy.
Now the simulation is made at off-critical temperature in the
ordered phase and the maximum Monte Carlo step is up to
$90$ or $35$. So the results can also be regarded as the values in the
thermodynamics limit within the present accuracy.

The estimated values of magnetization are plotted in Fig.~\ref{LOWLOGLOG}.
The exact solution of the spontaneous magnetization,
\begin{equation}
m_s = \lbrack 1-{1\over \sinh^4(2K)}\rbrack^{1/8},
\end{equation}
is used.
The asymptotic slopes of these curves correspond to the value of $\alpha$.

At $t=0$ where $m=1$, the values of $\ln [-\ln (m-m_s)]$ are
$0.647$, $0.819$, $0.943$ and $1.041$ for $K=0.47$, $0.49$, $0.51$ and
$0.53$, respectively, and these values correspond to $\ln t =-\infty$.
Therefore the tangents of the curves in Fig.~\ref{LOWLOGLOG} approach
the final value from smaller side and the curves are concave.

The value of $\alpha$ observed from the Fig.~\ref{LOWLOGLOG}
is clearly larger than $1/3$.
The results for $K=0.51$ and $0.53$ are consistent with $\alpha =1/2$.
For $K=0.47$ and $0.49$, the curved do not reach $\alpha =1/2$ behavior yet,
but the final tangent are already larger than $1/3$ which is observed in
Fig.~\ref{LOWK047}.

We cannot exclude the possibility that the curves continue to become
steeper for longer time and finally the simple exponential decay($\alpha =1$)
appears.
We can conclude that the conjecture $\alpha =1/3$ is not true for
the non-equilibrium relaxation of this model in the presently studied
time region.

\section{Three-dimensional Ising model}

The interface energy of the three-dimensional Ising model is studied in this
section.

\subsection{Monte Carlo simulation}

The interface energy is estimated by the difference of the energy in the
systems with and without a interface.
In this study, the different boundary conditions are used to realize
the constraint for the interface.
One lattice has periodic boundary conditions for all of the $x$, $y$ and $z$
directions.
In this case, the configuration without interface is dominant because the
present simulation is started from all up initial configuration
and we take lattice size to infinity at fixed $t$.
The other lattice has anti-periodic boundary condition in the $z$ direction
and periodic to the $x$ and $y$ direction.
This means that the lattice is pasted into torus with anti-ferromagnetic
bond with coupling constant $-K(<0)$ for $z$ direction.
In this case, a single interface
configuration is dominant because the all-up initial configuration
has one interface in the case of the anti-periodic lattice.

The lattice size is specified by two integers, $L$ and $H$, in the following.
We use the box of cubic lattice of the size of $L\times L\times H$
for $x$, $y$ and $z$ directions.
Odd $L$ and even $H$ are used for the purpose of vectorizing
the Monte Carlo dynamics.

The total energies of periodic and antiperiodic lattices are denoted by
$E_{\rm p}$ and $E_{\rm ap}$.
The interface energy per unit area denoted by $\epsilon_{s}$ is defined
by
\begin{equation}
\epsilon_s(K,L,H) = (E_{\rm ap} - E_{\rm p})/L^2.
\end{equation}

In the thermodynamic limit, the critical behavior of this
interface energy is expected to be
\begin{equation}
\epsilon_s(K) \equiv \epsilon_s(K,L=\infty, H=\infty )
  = {2\nu \sigma_0 \over K} (1-{K_c\over K})^{2\nu -1},
\end{equation}
where $\sigma_0$ denotes the critical amplitude of the surface tension and
the factor $1/K$ is added instead of $1/K_c$ because it
gives a better description of the present results and the critical behavior is
essentially the same.

The simulations listed in Table~\ref{SEMCS} were made on MONTE-4 of JAERI
and NEC SX3/11 of K\"oln University.
In each simulation 64 independent simulations were run simultaneously
using the independent-system coding technique\cite{IK88,IK90A,KIK92}
and the recycle algorithm for the random number\cite{IKO93}.
Half of these 64 systems have the periodic boundary condition and the
rest have the anti-periodic boundary condition.
All the simulations here are made in the ordered phase above the surface
roughening temperature.
The triplet $(A$, $B$, $C)$ in the MCS column of Table~\ref{SEMCS}
denotes the length of the simulation and the energy measurement timing.
{}From all-up initial configuration, $A$ MCS are made without
energy calculation.
The value of $A$ is selected to be twenty or more times larger than the
correlation time of the energy.
Then the energy is calculated $C$ times in every $B$ MCS and one estimate
of the interface energy is calculated using the estimates of the energy
of $64$ systems.
These operations are repeated five times independently.
If five run can be made within the CPU time limit of one
batch job, the initialization of the configuration is not made for
every run but the last configuration of the previous run is used as
the next initial configuration.
The obtained five estimates of the interface energy are used to estimate
the average, $\epsilon_s(K,L,H)$, and its error.

\subsection{Finite size effect}

For a fixed $L$, the interface energy will be independent of the height $H$
if $H$ is larger than the interface thickness which is known to increase
as $\log L$,
as far as there is no interface in the periodic lattice and there is only one
interface in the anti-periodic lattice.
Fig.~\ref{SEHDEP} shows the $H$ dependence of the estimates of
$\epsilon_s(K,L,H)$.
It is observed that $\epsilon_s(K,L,H)$ is independent of $H$
if $H$ is not very small.
The horizontal solid line and the broken lines show the estimate
of $\epsilon_s(K,L)\equiv \epsilon_s(K,L,H=\infty )$
and its $1\sigma$ error region, which is
estimated from the data $H\leq 20$, that is, all the data except two
small $H$ points shown in the figure.

For every $L$ at every $K$, lattices with several $H$ are
made.
The values of the interface energy were estimated from the
estimates for the $H$-independent region.
The region of $H$ used for this procedure
is given in Table~\ref{SEHAVE} with the values of $\epsilon_s$.

Now the extrapolation to $L\rightarrow\infty$ has to be made.
The finite size correction to $\epsilon_s(K)$ is expected to
be the order of $1/L^2$ from the capillary wave contribution,
that is,
\begin{equation}
\epsilon_s (K,L) = \epsilon_s (K, L=\infty ) + {a\over L^2},
\end{equation}
up to the first order correction.
The values of the interface energy in the thermodynamic limit
$\epsilon_s(K) = \epsilon_s(K,L=\infty )$ are estimated from the
estimates of $\epsilon_s (K,L)$ with this size dependence.
This $1/L^2$ correction is appropriate to our results.
One example of this extrapolation is shown in Fig.~\ref{SELDEP}.
The estimated values of $\epsilon_s(K)$ are also given in
Table~\ref{SEHAVE}.

\subsection{Critical behavior}

The critical exponent $\nu$ and the critical amplitude $\sigma_0$ are
estimated from our estimates of $\epsilon_s$.
The present estimates of $\epsilon_s$ are plotted in Fig.~\ref{SELOGPLOT}.

The value of the critical point $K_c$ has been estimated by several methods
and usually is in the region $0.22165 \sim 0.22166$\cite{BGHP92} and
this accuracy is enough precise for the present analysis.
So $K_c=0.221657$\cite{IS91,NI92} is used for the present analysis.

The tangent of the consecutive two points in Fig.~\ref{SELOGPLOT}
are shown in Fig.~\ref{SENUPLOT}
and the value of $2\nu -1$ at $K=K_c$ is estimated to be $0.250(5)$
from this figure.
This means that the estimate of $\nu$ is $0.6250\pm 0.0025$.
This values of $\nu$ is consistent with the result of a recent Monte Carlo
renormalization study\cite{BGHP92}. The previous estimates of $\nu$ are
also cited in Ref.~\cite{BGHP92}.

The value of $\sigma_0$ is estimated from Fig.~\ref{SES0PLOT} which
shows the values of
\begin{equation}
\sigma_0(\nu) = \epsilon_s(K)\cdot {K\over 2\nu } (1-{K_c\over K} )^{1-2\nu },
\end{equation}
from our estimates of $\epsilon_s(K)$ for several values of $\nu$.
{}From this figure and above estimate of $\nu$, we estimate that the
value of $\sigma_0$ is $1.42 \pm 0.04$.
The estimates in the previous works are shown in Table~\ref{SEPREV}
and are mostly confirmed by our more accurate data.

\section{Summary}

First the non-equilibrium relaxation of the two-dimensional
ferromagnetic Ising model was studied.
The critical relaxation exponent $z$ was estimated to be $2.165\pm 0.010$.
This value is smaller than the recent other estimations for this
value from the same non-equilibrium relaxation method
by M\"unkel et al\cite{MH93} and from the high-temperature expansion
study by Damman and Reger\cite{HTE92A}.
But the upward behavior of the two-dimensional data shown in
Fig.~1 of Ref.~\cite{MH93} can be interpreted as a sample fluctuation
and this might cause a overestimation.
The coefficients of the high-temperature expansion given in
Ref.~\cite{HTE92A} were reanalyzed by Adler\cite{JA93} and an estimate
consistent with the present result was reported.
The relaxation process in the ordered phase is stretched exponential.
The exponent is larger than $1/3$\cite{TNM88} and it is consistent with
$1/2$\cite{HF87}.

Secondly the critical exponent $\nu$ and the critical amplitude $\sigma_0$
are estimated to be $0.6250\pm 0.0025$ and $1.42\pm 0.04$, respectively.
The fully vectorized simulation with single spin updating dynamics
can efficiently study the interface. When we used only about 20 hours
of NEC SX3/11, the data were as accurate as those of Ref.~\cite{HP93}
from $10^3$ work station hours.

Finally the new Monte Carlo renormalization group method produced
reasonable estimate of $\beta/\nu$ with a quite small scale simulation.

The total number of updated spins are $3.19\times 10^{15}$.

\section*{Acknowledgment}

The author thanks D.~Stauffer for his valuable advice and
discussion.
This study is partially supported by SFB341.

\section*{Appendix~A}

A new Monte Carlo renormalization group method is applied to the
estimation of $\beta /\nu$ value of the three-dimensional Ising model.
The value of the magnetization of the scale transformed configuration
with scale $l=2$ majority-rule real-space scale transformation is estimated.
A $256^3$ lattice is simulated up to $20000$MCS from all-up initial
configuraion with an appropriate Monte Carlo dynamics.
Totally 45 independent simulations are made and therefore the total number
of the updated spin is $1.51\times 10^{13}$.

Following the eq.~(\ref{EQFRACT}), the values of $\lambda (n,t)$ defined by
\begin{equation}
\lambda (n,t)= {\log [M(n-1,t)/M(n,t)] \over \log l}
\end{equation}
are studied.
This $\lambda (n)$ is expected to be $d-\beta /\nu$, where $d$ is the lattice
dimensionality, in the limit of $t\rightarrow \infty$ if the infinitely
large lattice is studied.
The behavior of the $\lambda (n,t)$ is shown in Fig.~\ref{FRDIMBEHAV}.
{}From this figure, the value of $\lambda (2,t=\infty )$ is estimated to be
$2.4625(1)$.
In table~\ref{FRVALUE}, the estimated values of
$\lambda (n) \equiv \lambda (n,t=\infty )$
are shown for $256^3$ and $1024^3$ lattice.
The $1024^3$ lattice was simulated up to $5644$ MCS for one sample and
the number of the updated spin is $6.06 \times 10^{12}$.

When we extrapolate these estimates of $\lambda (n)$ are extrapolated to
$n\rightarrow \infty$ with
\begin{equation}
\lambda (n) = \lambda + c l^{-n\omega },
\end{equation}
the value of $\lambda$ behaves as shown in Fig.~\ref{FREXTR}.
The estimates for the $256^3$ lattice given in Table~\ref{FRVALUE} are
used for the extrapolation.
{}From the present preliminary data, the value of $\omega$ is estimated
to be $0.7(2)$ and the corresponding value of $\lambda$ is $2.492(8)$.
this means the value of $\beta /\nu = 0.508(8)$.
If the estimate $\omega = 0.80 \sim 0.85$\cite{BGHP92} is assumed,
$\lambda$ is estimated to be $2.487 \sim 2.490$ and $\beta /\nu
= 0.513 \sim 0.510$. If we assume $\omega = 1$, $\lambda$ and $\beta /\nu$
are $2.486$ and $0.514$, respectively.

\section*{Appendix~B}
Two routines of IS2DVP code which are specific to the two-dimensional Ising
model are shown.
BFL2DF generates the Boltzmann factor tables and SU2DSK updates the spin
configuration.

{
\begin{verbatim}
C**************************************************************
C IS2DVP: Monte Carlo simulation of two-dimensional Ising Model
C
C Boltzmann-factor table preparation routine
C     METROPOLIS-TYPE TRANSITION PROBABILITY
C     1992.09.19. VERSION 1.00 BY NOBUYASU ITO
C
C TK(64) R*8 : Give a inverse temperature of 64 systems
C IX1(0:IRLST), IX2(0:IRLST) I*8 : The BFT is generated in them.
C     IRLST is maxinum number of the random integer.
C**************************************************************
      SUBROUTINE BFL2DF(TK,IX1,IX2,IRLST)
      PARAMETER(IWIDTH=64)
      DIMENSION TK(IWIDTH)
      REAL*8 TK
      DIMENSION IX1(0:IRLST),IX2(0:IRLST)
      DIMENSION IK1(IWIDTH),IK2(IWIDTH),I2P(IWIDTH)
      REAL*8 TNORM
C
      TNORM=DFLOAT(IRLST)
      DO 10 I=1,IWIDTH
        IK1(I)=IDINT(TNORM*DEXP(-8.0D0*TK(I)))
        IK2(I)=IDINT(TNORM*DEXP(-4.0D0*TK(I)))
   10 CONTINUE
      DO 30 I=1,IWIDTH
        I2P(I)=ISHFT(1,(IWIDTH-I))
   30 CONTINUE
      DO 60 I=0,IRLST
      IX1(I)=0
      IX2(I)=0
   60 CONTINUE
*VDIR NODEP
      DO 50 I=1,IWIDTH
      DO 40 J=0,IRLST
        IXT1=0
        IXT2=0
        IF(J.LE.IK1(I))IXT1=I2P(I)
        IF(J.LE.IK2(I))IXT2=I2P(I)
        I2=IEOR(IXT1,IXT2)
        I1=IAND(IXT1,IXT2)
        IX2(J)=IOR(IX2(J),I2)
        IX1(J)=IOR(IX1(J),I1)
   40 CONTINUE
   50 CONTINUE
      RETURN
      END
\end{verbatim}

\begin{verbatim}
C**************************************************************
C IS2DVP: Monte Carlo simulation of two-dimensional Ising Model
C
C Spin configuration updating routine for
C  ferromagnetic Ising model with Metropolis-type transition
C  probabilities
C BFL2DF routine should be used for BFT preparation
C     1992.09.19. VERSION 1.00 BY NOBUYASU ITO
C
C  ISTEP I*8 : number of Monte Carlo sweeps
C  L1, L2 I*8: lattice size
C  IS((-L1+1):(L1*(L2+1)) I*8: spin configuration
C  IRD(L1*L2) I*8: work area
C  IRLST  I*8: Maxinum number of random integer
C  IX1(0:IRLST), IX2(0:IRLST) I*8: Boltzmann factor tables(BFT)
C**************************************************************
      SUBROUTINE SU2DSK(ISTEP,L1,L2,IS,IRD,IRLST,IX1,IX2)
      DIMENSION IS((-L1+1):(L1*(L2+1)))
      DIMENSION IRD(L1*L2)
      DIMENSION IX1(0:IRLST),IX2(0:IRLST)
      NSYS=L1*L2
      LS=L1
      DO 10 IMCS=1,ISTEP
      CALL RNDO2I(NSYS,IRD)
      IFIRST=1
*VDIR NODEP
   40 DO 20 I=-LS+1,0
   20 IS(I)=IS(I+NSYS)
*VDIR NODEP
      DO 30 I=NSYS+1,NSYS+LS
   30 IS(I)=IS(I-NSYS)
*VDIR NODEP
      DO 50 IJ=IFIRST,NSYS,2
        I0=IS(IJ)
        I1=IS(IJ+1)
        I2=IS(IJ-1)
        I3=IS(IJ+L1)
        K1=IEOR(I1,I2)
        K2=IAND(I1,I2)
        J2=IEOR(K1,I3)
        K3=IAND(K1,I3)
        J1=IOR(K2,K3)
        J1=IEOR(I0,J1)
        J2=IEOR(I0,J2)
        I4=IS(IJ-L1)
        I4=IEOR(I4,I0)
        IRT=IRD(IJ)
        IX1T=IX1(IRT)
        IX2T=IX2(IRT)
        K2=IEOR(I4,IX2T)
        K1T=IAND(I4,IX2T)
        K1=IOR(K1T,IX1T)
        ID=IOR(J1,K1)
        K4=IAND(J2,K2)
        ID=IOR(ID,K4)
        IS(IJ)=IEOR(IS(IJ),ID)
   50 CONTINUE
      IF(IFIRST.EQ.1)THEN
        IFIRST=2
        GOTO 40
      END IF
   10 CONTINUE
      RETURN
      END
\end{verbatim}
}




\begin{figure}[p]
\caption{The size-dependences of $m(t)$ at (a)$t=10$, (b)$100$,
(c)$400$ and (d)$500$ are shown.
}
\label{CRITSIZEDEP}
\end{figure}

\begin{figure}[p]
\caption{The estimated values of m(t) for $L=1500$ lattice up to
$400$MCS are shown in logarithmic scale.
They are regarded as those in the thermodynamics limit within
the present accuracy.
}
\label{CRITMAGNET}
\end{figure}

\begin{figure}[p]
\caption{
The value of the averaged local exponent $z_{\mbox{ave.}}^{25}$ is shown.
The magnetization obtained for $L=1500$ lattice is used.
}
\label{CRITRATIO}
\end{figure}

\begin{figure}[p]
\caption{
The estimates of $z$ from the least-square fitting are shown.
$\tau (t,50)$ denotes the mean time of the data, that is, $t+(50-1)/2$.
}
\label{CRITFITTING}
\end{figure}

\begin{figure}[p]
\caption{
The values of $m(t)$ is plotted assuming the logarithmic correction.
}
\label{CRITLOGLOG}
\end{figure}

\begin{figure}[p]
\caption{
The estimates for $z$ are plotted.
The points marked with $\bullet$ and $\circ$ show simulational
and other results, respectively.
The horizontal axis shows the publication year except the recent
preprints.
}
\label{CRITZPLOT}
\end{figure}

\begin{figure}[p]
\caption{
The non-equilibrium relaxation curves of magnetization are shown.
The point marked by $\bullet$, $\circ$, $\triangle$ and $\diamond$
correspond to $K=0.47$, $0.49$, $0.51$ and $0.53$, respectively.
The tangent of the sold line is $\alpha =1/2$.
Those of broken lines are $\alpha =1/3$(gentler one) and $1$(steeper one).
}
\label{LOWLOGLOG}
\end{figure}

\begin{figure}[p]
\caption{
The magnified figure of the magnetization at K=0.47.
The solid line has the slope of 1/3. It is observed that
the exponent $\alpha$ is already larger than $1/3$.
}
\label{LOWK047}
\end{figure}

\begin{figure}[p]
\caption{
The height dependence of the interface energy is shown.
The estimates of $\epsilon_s(K,L,H)$ for $L=21$ are plotted as an example.
}
\label{SEHDEP}
\end{figure}

\begin{figure}[p]
\caption{
The $L$ dependence of the interface energy is shown.
The estimates of $\epsilon_s(K,L)$ at $K=0.24$ are plotted as an example.
}
\label{SELDEP}
\end{figure}

\begin{figure}[p]
\caption{
The critical behavior of the interface energy is shown.
$K\epsilon_s(K)$ is plotted versus $K-K_c$ in the logarithmic scale.
}
\label{SELOGPLOT}
\end{figure}

\begin{figure}[p]
\caption{
The values of the exponent $2\nu -1$ from  the consecutive two points
are shown.
}
\label{SENUPLOT}
\end{figure}

\begin{figure}[p]
\caption{
The values of $\sigma_0(\nu)$ are plotted.
$*$, $\triangle$, $\bullet$, $\diamond$ and $\circ$ show the estimates
for $\nu = 0.6300$, $0.6275$, $0.6225$ and $0.6200$, respectively.
}
\label{SES0PLOT}
\end{figure}


\begin{figure}[p]
\caption{
The behavior of the $\lambda (2, t)$ is shown.
Every point denotes an averaged value of $\lambda (2,t)$ over successive
$1000$ MCS.
}
\label{FRDIMBEHAV}
\end{figure}

\begin{figure}[p]
\caption{
The extrapolated values of $\lambda (n)$ to $n\rightarrow \infty$
assuming the value of $\omega$ are plotted.
}
\label{FREXTR}
\end{figure}

\pagebreak

\begin{table}[p]
\begin{center}
\begin{tabular}{cccc}\hline
machine&$N_{\rm job}$& performance& \\
       &              &    U & U+M \\\hline
one processor of& 0 & $1.31$ & $0.77$ \\
MONTE-4         & 1 & $1.1\sim 1.2$ & $0.6\sim 0.7$ \\
& 2 & $0.9\sim 1.0$ & $0.5$ \\
& 3 & $0.7 \sim 0.9 $ & $0.4\sim 0.5$ \\\hline
SX3/11          & $-$ & $1.05$ & $0.63$ \\\hline
\end{tabular}
\end{center}
\caption{
The performance of the simulation code, IS2DVP, is shown.
The simulation speed in the case of $1501\times 1500$ lattice is shown.
The unit of the figures is GUPS, that is, giga($10^9$) update per second.
The figures in "U" and "U$+$M" denote the speed of the spin updating
procesures only and of the spin updating procesure with
magnetization calculation in every Monte Carlo sweep.
The number $N_{\rm job}$ in MONTE-4 shows the number of concurrent jobs
on other processors of MONTE-4.
The concurrent jobs are highly vectorized with dense memory access.
Therefore the speed in $N_{\rm job}\geq 1$ case shows the worst case.
}
\label{PFIS2DVP}
\end{table}

\begin{table}[p]
\begin{center}
\begin{tabular}{ccc}\hline
$L$ & $N_{\mbox{sample}}$ &updated spins \\\hline
$1000$ & $1.54\times 10^5$ & $7.69\times 10^{13}$ \\
$1500$ & $1.18\times 10^6$ & $1.33\times 10^{15}$ \\
$2000$ & $6.14\times 10^4$ & $1.23\times 10^{14}$ \\
$3000$ & $1.54\times 10^4$ & $6.91\times 10^{13}$ \\\hline
\end{tabular}
\end{center}
\caption{
The simulations made for the non-equilibrium critical relaxation
study are listed. $N_{\mbox{sample}}$ denotes the number of independent runs.
The simulation up to $500$MCS is made for every sample from
all-up initial condition.
}
\label{KCSIMUL}
\end{table}

\begin{table}[p]
\begin{center}
\begin{tabular}{cccc}\hline
year&method&$z$&reference\\\hline
$1968$&   &$1.75$ & \cite{OTHER68A}\\
$1969$&HTE&$2.00 \pm 0.05$ & \cite{HTE69A}\\
$1971$&HTE&$2$&\cite{HTE71A}\\
$1972$&RG &$2.0$&\cite{RG72A}\\
$1973$&MC &$1.90\pm 0.10$&\cite{MC73A}\\
$1975$&RG &$1.82$&\cite{RG75A}\\
$1976$&MC &$2.30\pm 0.30$&\cite{MC76A}\\
$1976$&HTE &$2.125\pm 0.01$&\cite{HTE76A}\\
$1976$&MCRG &$1.4\pm 0.4$&\cite{MCRG76A}\\
$1978$&RG &$2.2$&\cite{RG78A}\\
$1978$&RG &$2.19$&\cite{RG78B}\\
$1978$&RG &$1.7$&\cite{RG78C}\\
$1978$&RG &$1.67$&\cite{RG78D}\\
$1979$&MC+FSS &$2.0\pm 0.1$&\cite{MC79A}\\
$1979$&RG &$2.09$&\cite{RG79A}\\
$1979$&RG &$1.96$&\cite{RG79B}\\
$1979$&RG &$2.064$(energy)&\cite{RG79C}\\
$1979$&RG &$1.819$(magnetization)&\cite{RG79C}\\
$1979$&MCRG &$1.85\pm 0.15$&\cite{MCRG79A}\\
$1980$&TM+FSS &$1.99$&\cite{TM80A}\\
$1980$&RG &$2.22$&\cite{RG80A}\\
$1980$&RG &$1.819$&\cite{RG80B}\\
$1980$&RG &$2.2$&\cite{RG80C}\\
$1980$&RG &$1.791$&\cite{RG80D}\\
$1981$&MC &$1.9$&\cite{MC81A}\\
$1981$&MCRG &$2.22\pm 0.13$&\cite{MCRG81A}\\
$1981$&RG &$2.126$&\cite{RG81A}\\
$1982$&MC+NER&$2.1$&\cite{MC82A}\\
$1982$&MC+FSS &$2.0$&\cite{MC82B}\\
$1982$&MC &$2.2\pm 0.1$&\cite{MC82C}\\
$1982$&MCRG &$2.23$&\cite{MCRG82A}\\
$1982$&MCRG &$2.08$&\cite{MCRG82B}\\
\end{tabular}
\end{center}
\end{table}

\begin{table}
\begin{center}
\begin{tabular}{cccc}
$1983$&MCRG&$2.12$&\cite{MCRG83A}\\
$1983$&RG &$2.2$&\cite{RG83A}\\
$1984$&RG,HTE &$2.32$&\cite{RGHTE84A}\\
$1984$&MCRG+NER &$2.14\pm 0.02$&\cite{MCRG84A}\\
$1984$& &$2$&\cite{OTHER84A}\\
$1985$&RG &$1.73$&\cite{RG85A}\\
$1985$&RG &$1.856$&\cite{RG85B}\\
$1985$&MC &$2.2$&\cite{MC85A}\\
$1985$&MCRG &$2.13\pm 0.03$&\cite{MCRG85A}\\
$1986$&Q2R &$2.1\pm 0.1$&\cite{Q2R86A}\\
$1986$&MCRG &$2.19\pm 0.05$&\cite{MCRG86A}\\
$1987$&MC+NER&$2.16\pm 0.05$&\cite{MC87A}\\
$1987$&MC+FSS &$2.132\pm 0.008$&\cite{MC87B}\\
$1987$&MC+FSS &$2.17\pm 0.06$&\cite{MC87C}\\
$1988$&Q2R &$2.16$&\cite{Q2R88A}\\
$1988$&MC&$2.22$&\cite{Q2R88A}\\
$1988$&MC+NER &$2.076\pm 0.005$&\cite{MC88A}\\
$1990$&HTE &$2.34\pm 0.03$&\cite{HTE90A}\\
$1992$&MC+NER&$2.2$&\cite{MC92A}\\
$1992$&HTE &$2.183\pm 0.005$&\cite{HTE92A}\\
$1993$&HTE &$2.21\pm 0.005$&\cite{HTE93B}\\
$1993$&MC+NER&$2.21\pm 0.03$&\cite{MH93}\\
$1993$&HTE&$2.165\pm 0.015$&\cite{JA93}\\\hline
$1993$&MC+NER&$2.165\pm 0.010$&present study\\\hline
\end{tabular}
\end{center}
\caption{
The estimates of $z$ are listed.
The HTE, RG, MC, FSS and NER in this table are the abbreviations of
high-temperature expansion, renormalization group, Monte Carlo,
finite-size scaling and non-equilibrium relaxation, respectively.
}
\label{CRITZEST}
\end{table}

\begin{table}[p]
\begin{center}
\begin{tabular}{ccccc}\hline
$K$&$L$&$T_{\mbox{max.}}$&$N_{\mbox{sample}}$&updated spins\\\hline
$0.47$ & $1500$ & $90$ & $1.02\times 10^6$ & $2.07\times 10^{14}$ \\
$0.49$ & $2000$ & $35$ & $3.58\times 10^5$ & $5.02\times 10^{13}$ \\
$0.51$ & $2000$ & $35$ & $3.58\times 10^5$ & $5.02\times 10^{13}$ \\
$0.53$ & $2000$ & $35$ & $1.54\times 10^5$ & $2.15\times 10^{13}$ \\\hline
\end{tabular}
\end{center}
\caption{
The simulations made for the relaxation study in the ordered phase.
$K$ and $T_{\mbox{max.}}$ show the inverse temperature and the Monte
Carlo steps for each run.
}
\label{LOWMCS}
\end{table}

\begin{table}[p]
\begin{center}
\begin{tabular}{rlll}\hline
$L$ &$K$ & $H$ & MCS \\\hline
$11$ & $0.23$, $0.24$, $0.25$ & $6$, $10$, $16$, $20$, $26$, $30$
     & $(1$K, $10$, $1$K$)$ \\\hline

$15$ & $0.23$, $0.24$, $0.25$ & $6$, $10$, $16$, $20$, $26$, $30$, $36$,
       $40$, $46$
     & $(1$K, $10$, $1$K$)$ \\\hline

$21$ & $0.23$, $0.24$, $0.25$ & $10$, $16$, $20$, $30$, $40$, $50$, $60$
     & $(2$K, $10$, $2$K$)$ \\
     &$0.23$&$80$, $100$, $150$, $200$, $250$, $300$, $350$, $400$&
       $(2$K, $10$, $2$K$)$ \\
     &$0.225$, $0.2275$&$50$, $74$, $100$, $124$, $150$, $174$, $200$, $250$
     & $(10$K, $10^2$, $1$K$)$ \\\hline

$25$ & $0.23$, $0.24$, $0.25$ & $10$, $20$, $30$, $40$, $50$, $60$, $70$, $80$
     & $(2$K, $10$, $1$K$)$ \\\hline

$31$ & $0.23$, $0.24$, $0.25$ & $10$, $20$, $30$, $40$, $50$, $60$, $70$,
       $80$, $90$  & $(2$K, $20$, $1$K$)$ \\
     & $0.235$, $0.245$ & $50$, $60$, $70$, $80$, $90$
     & $(2$K, $10$, $2$K$)$ \\
     &$0.225$, $0.2275$&$50$, $74$, $100$, $124$, $150$
     & $(10$K, $10^2$, $1$K$)$ \\\hline

$35$ & $0.23$, $0.24$, $0.25$ & $20$, $40$, $60$, $80$, $100$, $120$
     & $(2$K, $20$, $1$K$)$ \\
     & $0.235$, $0.245$ & $50$, $60$, $70$, $80$, $90$
     & $(2$K, $10$, $2$K$)$ \\\hline

$41$ & $0.23$, $0.24$, $0.25$ & $20$, $40$, $60$, $80$, $100$, $120$
     & $(2$K, $10$, $1$K$)$ \\
     & $0.235$, $0.245$ & $50$, $60$, $70$, $80$, $90$
     & $(2$K, $10$, $2$K$)$ \\
     &$0.225$, $0.2275$&$100$, $124$, $150$
     & $(20$K, $200$, $1$K$)$ \\\hline

$45$ & $0.235$, $0.245$ & $50$, $60$, $70$, $80$, $90$
     & $(2$K, $10$, $2$K$)$ \\\hline

$51$ & $0.23$, $0.24$, $0.25$ & $26$, $50$, $76$, $100$, $126$, $150$, $176$,
       $200$     & $(2$K, $10$, $2$K$)$ \\
     & $0.235$, $0.245$ & $50$, $60$, $70$, $80$, $90$
     & $(2$K, $10$, $2$K$)$ \\
     &$0.225$, $0.2275$&$100$, $124$, $150$
     & $(20$K, $200$, $1$K$)$ \\\hline

$61$ & $0.23$, $0.24$, $0.25$ & $30$, $60$, $90$, $120$, $150$, $180$, $210$
     & $(2$K, $10$, $2$K$)$ \\
     &$0.225$, $0.2275$&$100$, $124$, $150$&$(20$K, $200$, $500)$ \\
     & $0.235$, $0.245$ & $50$, $60$, $70$, $80$, $90$
     & $(2$K, $10$, $2$K$)$ \\\hline

$71$ &$0.225$, $0.2275$&$100$, $124$, $150$&$(20$K, $200$, $500)$ \\\hline

$81$ & $0.23$, $0.24$, $0.25$ & $40$, $60$, $80$, $100$, $120$
     & $(2$K, $10$, $2$K $)$ \\
     & $0.235$, $0.245$ & $50$, $60$, $70$, $80$, $90$
     & $(2$K, $10$, $2$K$)$ \\\hline

$101$ &$0.225$&$100$&$(20$K, $200$, $400)$ \\
 &$0.225$&$124$&$(20$K, $200$, $300)$ \\
 &$0.225$&$150$&$(20$K, $200$, $250)$ \\
 &$0.23$, $0.24$, $0.25$&$50$, $60$, $70$, $80$, $90$, $100$&
  $(2$K, $10$, $2$K$)$ \\\hline

$151$&$0.23$, $0.24$, $0.25$&$50$, $60$, $70$, $80$, $90$ &
  $(2$K, $10$, $2$K$)$ \\\hline

$171$ &$0.23$&$50$, $70$, $90$&$(2$K, $10$, $2$K$)$ \\\hline




\end{tabular}
\end{center}
\caption{
The simulations made for the interface study are listed.
The three numbers in MCS column show the MCS schedules and
their meaning is explained in the text($1$K $= 1000$).
Totally $1.26\times 10^{15}$ spins are updated.
}

\label{SEMCS}
\end{table}

\begin{table}[p]
\begin{center}
\begin{tabular}{crccl}\hline
$K$ & $L$ & $H$ & $N_{H}$ & $\epsilon_{s}$ \\\hline
$0.225$ & $ 31$&$ 50$--$150$&$ 5$&$2.5592(32)$\\
  &$ 41$&$100$--$150$&$ 3$&$2.6342(56)$\\
  &$ 51$&$100$--$150$&$ 3$&$2.6741(32)$\\
  &$ 61$&$100$--$150$&$ 3$&$2.6977(43)$\\
  &$ 71$&$100$--$150$&$ 3$&$2.7036(39)$\\
  &$101$&$100$--$150$&$ 3$&$2.7234(25)$\\\cline{2-5}
  &$\infty$ & -- & -- & $2.74(1)$ \\\hline
$0.2275$  &$ 21$&$ 50$--$250$&$ 8$&$2.8858(56)$\\
  &$ 31$&$ 50$--$150$&$ 5$&$2.9938(40)$\\
  &$ 41$&$100$--$150$&$ 3$&$3.0365(51)$\\
  &$ 51$&$100$--$150$&$ 3$&$3.0578(37)$\\
  &$ 61$&$100$--$150$&$ 3$&$3.0725(49)$\\
  &$ 71$&$100$--$150$&$ 3$&$3.0832(32)$\\\cline{2-5}
  &$\infty$ & -- & -- & $3.10(1)$ \\\hline
$0.23$  &$ 11$&$ 16$--$ 30$&$ 4$&$2.7649(89)$\\
  &$ 15$&$ 16$--$ 46$&$ 7$&$3.0443(30)$\\
  &$ 21$&$ 30$--$400$&$12$&$3.1726(20)$\\
  &$ 25$&$ 30$--$ 80$&$ 6$&$3.2200(28)$\\
  &$ 31$&$ 30$--$ 90$&$ 7$&$3.2608(17)$\\
  &$ 35$&$ 40$--$120$&$ 5$&$3.2790(35)$\\
  &$ 41$&$ 40$--$120$&$ 5$&$3.2904(22)$\\
  &$ 51$&$ 76$--$200$&$ 6$&$3.3034(13)$\\
  &$ 61$&$ 60$--$150$&$ 4$&$3.3118(14)$\\
  &$ 81$&$ 40$--$120$&$ 5$&$3.32151(95)$\\
  &$101$&$ 50$--$100$&$ 6$&$3.32622(58)$\\
  &$119$&$ 50$--$ 80$&$ 4$&$3.32875(66)$\\
  &$151$&$ 50$--$ 90$&$ 5$&$3.33083(40)$\\
  &$171$&$ 50$--$ 90$&$ 3$&$3.33128(57)$\\\cline{2-5}
  &$\infty$ & -- & -- & $3.335(3)$ \\\hline
\end{tabular}
\end{center}
\end{table}

\begin{table}[p]
\begin{center}
\begin{tabular}{crccl}\hline
$K$ & $L$ & $H$ & $N_{H}$ & $\epsilon_{s}$ \\\hline
$0.235$  &$ 31$&$ 50$--$ 90$&$ 5$&$3.5859(17)$\\
  &$ 35$&$ 50$--$ 90$&$ 5$&$3.5939(11)$\\
  &$ 41$&$ 50$--$ 90$&$ 5$&$3.6029(18)$\\
  &$ 45$&$ 50$--$ 90$&$ 5$&$3.6073(13)$\\
  &$ 51$&$ 50$--$ 90$&$ 5$&$3.6132(14)$\\
  &$ 61$&$ 50$--$ 90$&$ 5$&$3.62088(83)$\\
  &$ 81$&$ 50$--$ 90$&$ 5$&$3.62346(67)$\\\cline{2-5}
  &$\infty$ & -- & -- & $3.630(5)$ \\\hline
$0.24$  &$ 11$&$ 16$--$ 30$&$ 4$&$3.5778(22)$\\
  &$ 15$&$ 16$--$ 46$&$ 7$&$3.6852(34)$\\
  &$ 21$&$ 20$--$400$&$13$&$3.7447(14)$\\
  &$ 25$&$ 30$--$ 80$&$ 6$&$3.7616(26)$\\
  &$ 31$&$ 30$--$ 90$&$ 7$&$3.7829(19)$\\
  &$ 35$&$ 20$--$120$&$ 6$&$3.7875(15)$\\
  &$ 41$&$ 20$--$120$&$ 6$&$3.7943(13)$\\
  &$ 51$&$ 26$--$200$&$ 8$&$3.79888(24)$\\
  &$ 61$&$ 30$--$210$&$ 7$&$3.80299(95)$\\
  &$ 81$&$ 40$--$120$&$ 5$&$3.80849(81)$\\
  &$101$&$ 50$--$100$&$ 6$&$3.81055(57)$\\
  &$151$&$ 50$--$ 90$&$ 5$&$3.81181(32)$\\\cline{2-5}
  &$\infty$ & -- & -- & $3.813(2)$ \\\hline
\end{tabular}
\end{center}
\end{table}

\begin{table}[p]
\begin{center}
\begin{tabular}{crccl}\hline
$K$ & $L$ & $H$ & $N_{H}$ & $\epsilon_{s}$ \\\hline
$0.245$  &$ 31$&$ 50$--$ 90$&$ 5$&$3.9101(16)$\\
  &$ 35$&$ 50$--$ 90$&$ 5$&$3.9141(13)$\\
  &$ 41$&$ 50$--$ 90$&$ 5$&$3.9181(10)$\\
  &$ 45$&$ 50$--$ 90$&$ 5$&$3.9202(16)$\\
  &$ 51$&$ 50$--$ 90$&$ 5$&$3.92080(67)$\\
  &$ 61$&$ 50$--$ 90$&$ 5$&$3.92691(61)$\\
  &$ 81$&$ 50$--$ 90$&$ 5$&$3.92687(63)$\\\cline{2-5}
  &$\infty$ & -- & -- & $3.930(2)$ \\\hline
$0.25$  &$ 11$&$ 16$--$ 30$&$ 4$&$3.8552(26)$\\
  &$ 15$&$ 16$--$ 46$&$ 7$&$3.9297(21)$\\
  &$ 21$&$ 20$--$400$&$13$&$3.96912(78)$\\
  &$ 25$&$ 20$--$ 80$&$ 7$&$3.9839(16)$\\
  &$ 31$&$ 30$--$ 90$&$ 7$&$3.9906(15)$\\
  &$ 35$&$ 20$--$120$&$ 6$&$3.9929(12)$\\
  &$ 41$&$ 20$--$120$&$ 6$&$4.0028(11)$\\
  &$ 51$&$ 26$--$200$&$ 8$&$4.00478(52)$\\
  &$ 61$&$ 30$--$210$&$ 7$&$4.00704(83)$\\
  &$ 81$&$ 40$--$120$&$ 5$&$4.00900(50)$\\
  &$101$&$ 50$--$100$&$ 6$&$4.01017(32)$\\
  &$151$&$ 50$--$ 90$&$ 5$&$4.01138(20)$\\\cline{2-5}
  &$\infty$ & -- & -- & $4.012(1)$ \\\hline
\end{tabular}
\end{center}
\caption{
The estimated values of $\epsilon_s$ for some values of $K$ and $L$ are shown.
The column $H$ shows the height region over which the average is calculated.
$N_H$ is the number of different $H$ used in the averaging operation.
The values of $\epsilon_s$ in $L=\infty$ denotes the estrapolated
values of $\epsilon_s(K,L)$ to $L=\infty$ and they are the estimates
for the interface energy in the thermodynamics limit.
}
\label{SEHAVE}
\end{table}

\begin{table}[p]
\begin{center}
\begin{tabular}{ccc}\hline
year & $\sigma_0$ & reference \\\hline
1982 & $1.05 \pm 0.05$ & \cite{KB82} \\
1985 & $1.2 \pm 0.1$ & \cite{MJ85} \\
1988 & $1.58 \pm 0.05$ & \cite{KKM88}\\
1993 & $1.5\pm 0.1$ & \cite{MH92}\\
1993 & $1.25 \sim 1.68$ & \cite{KM92} \\
1993 & $1.52\pm 0.05 $ & \cite{BHN92} \\
1993 & $1.221 \sim 1.492 $& \cite{HP93}\\
1993 & $1.92\pm 0.15$ & \cite{GPRS93} \\\hline
1993 & $1.42\pm 0.04$ & present study \\\hline
\end{tabular}
\end{center}
\caption{
The estimates of $\sigma_0$ are shown with the year of publication.
}
\label{SEPREV}
\end{table}

\begin{table}[p]
\begin{center}
\begin{tabular}{ccc}\hline
$n$ & $\lambda (n), 256^3$ & $\lambda (n), 1024^3$ \\\hline
$2$ & $2.4625(1)$ & $2.4624(2)$ \\
$3$ & $2.4738(5)$ & $2.4735(5)$ \\
$4$ & $2.481(1)$ & $2.480(2)$ \\
$5$ & $2.485(3)$ & $2.485(5)$ \\\hline
\end{tabular}
\end{center}
\caption{
The estimates of $\lambda (n)$ are shown.
The estimates from $256^3$ and $1024^3$ lattice coincide within
the present accuracy of the estimation.
}
\label{FRVALUE}
\end{table}

\end{document}